# UNAGI: A Conceptual Electrodynamic Tethered Spacecraft Demonstration Mission for Propellantless Landing on Io

Spencer Devins[1], Svitlana Kuklenko[2], Abhik Das[3], Cora Abrams-McCabe[3], Holly Herman[3], Madison Schooley[3], Matvey Zubkov[3], Sebastian Colussi[3], Dmitry Kuklenko[3], Abeljohn Burke-Raymond[3], Rae Chauvaux[3], Jonah Martinez[3], Mika Villanueva[3], Jorell Sakamoto[3], Alejandro Municio[3], Eduardo Duran[3], Elbert Chang[3], Advik Jain[3], Satvik Magganmane[3], Marcus Gonsalves[3], Antonio Corona[4], Anishka Chauhan[5], Zephra Blake[6]

*San Jose State University, San Jose, California, 95112, United States of America*

**This paper presents UNAGI, a novel spacecraft mission developed by Spartan Space Systems - a student engineering team at San Jose State University - aiming for the first controlled landing on Jupiter's volcanically active moon, Io. Inspired by the Japanese freshwater eel's agility, UNAGI employs an electrodynamic tether system that interacts with Jupiter's magnetic field to generate Lorentz forces, enabling dynamic modulation of the spacecraft's velocity without the need for traditional propellant. This system allows the spacecraft to precisely match Io's orbital velocity, akin to a surfer riding a wave. By optimizing tether current and orientation, UNAGI transitions from an outer-Jovian insertion orbit to an orbit around Io, achieving a controlled landing for on-surface science operations. The landing is strategically timed, leveraging the gravitational influence of Jupiter and its moons to decelerate descent and mitigate impact forces.**

**UNAGI leverages technologies from missions like Juno, JUICE, and Europa Clipper to ensure reliable operation in the challenging Jovian environment. The mission follows a phased approach, including cruise, orbital insertion, tether deployment, Jovian orbital operations, and final descent, with extensive risk mitigation through simulations and experimental validation. The concept supports rideshare integration on future NASA or ESA missions or a dedicated launch, emphasizing cost-effectiveness and adaptability. UNAGI carries a rich scientific payload, including infrared spectrometers, magnetometers, seismometers, and chromatographs, designed to probe Io's interior dynamics, volcanic activity, tectonic deformation, and magnetospheric interactions. This innovative mission promises to advance understanding of Io's geophysical processes and Jupiter system dynamics, setting a new standard for deep-space exploration of extreme environments.**

**Currently, the mission is in the preliminary spacecraft architecture phase, with ongoing development through detailed simulation and rigorous testing.**

[1] Undergraduate Student and Corresponding Author, Aerospace Engineering, AIAA Student member
[2] Graduate Student, Aerospace Engineering, AIAA Student member
[3] Undergraduate Student, Aerospace Engineering, AIAA Student member
[4] Undergraduate Student, Electrical Engineering, AIAA Student member
[5] Undergraduate Student, Computer Science, AIAA Student member
[6] Undergraduate Student, Political Science, AIAA Student member

**Nomenclature**

| | | |
|---|---|---|
| B | = | magnetic field strength |
| F | = | Lorentz force on tether |
| L | = | length of tether |
| r | = | radius of tether cross section |
| ε | = | induced voltage |
| η | = | resistivity of the tether material |
| θ | = | angle between the current vector and magnetic field vector |
| ν | = | translational velocity of the current-carrying wire through the magnetic field |

## I. Introduction

Deep-space exploration of Jupiter's moons presents a formidable challenge, particularly when targeting Io, the innermost Galilean satellite known for its intense volcanic activity and harsh radiation environment. Traditional chemical propulsion systems face significant mass limitations in such an extreme setting far from Earth, driving the need for innovative, propellantless approaches. This paper introduces UNAGI, a novel spacecraft mission concept developed by Spartan Space Systems at San Jose State University, which leverages electrodynamic tether (EDT) propulsion to achieve a controlled landing on Io.

Electrodynamic tethers offer a unique solution to planetary landings by harnessing the Lorentz force generated through the interaction between a long, conductive tether and Jupiter's intense magnetic field. Inspired by the agile movements of the Japanese freshwater eel, UNAGI's EDT system modulates spacecraft velocity without relying on conventional propellant. This capability not only provides precise trajectory corrections but also enables the spacecraft to synchronize its orbit with Io, facilitating a safe direct-descent and soft landing.

Building on decades of research—from NASA's Tethered Satellite System experiments to more recent small-scale tether demonstrations—this work outlines a comprehensive mission architecture that integrates state-of-the-art spacecraft design with advanced tether control strategies. The UNAGI mission is structured in phased operations, beginning with cruise and Jupiter orbit insertion, progressing through tether deployment and orbital adjustments, and culminating in a precision landing maneuver on Io. In parallel, the spacecraft carries a scientifically rich payload aimed at addressing critical decadal survey questions related to Io's interior dynamics and volcanic processes. Through this multidisciplinary approach, the UNAGI mission seeks not only to advance propulsion technology but also to significantly enhance our understanding of one of the solar system's most dynamic and intriguing bodies.

## II. Past Work and State of the Art

Electrodynamic tether propulsion has been explored as a promising alternative means of spacecraft maneuvering by harnessing the Lorentz force generated when a long, conductive tether interacts with a planetary magnetic field. Key experimental missions in this field include NASA's Tethered Satellite System-1 (TSS-1)[1] in 1992 aboard STS-46 and its follow-up, TSS-1R in 1996[2] on STS-75, both of which provided critical insights into tether dynamics, current collection, and thrust generation. The early 2000s saw the proposal of the ProSEDS[3] (Propulsive Small Expendable Deployer System) concept, designed to demonstrate tether-based propulsion and assist in space debris mitigation, although it did not fly. In recent years, smaller-scale and commercial initiatives have further advanced tether technology: ESTCube-1[4], a CubeSat launched on May 7, 2013, attempted a 10-meter tether deployment in low Earth orbit, with its successor ESTCube-2[5] failing to deploy from the ESA Vega upper stage in 2023. Complementing these efforts, the MiTEE spacecraft[6] developed by the University of Michigan has contributed valuable in-orbit insights into tether behavior, while the work of Brian Gilchrist has significantly advanced both the theoretical and practical frameworks underpinning electrodynamic tether systems.

Io, Jupiter's innermost Galilean moon, has long captivated planetary scientists, with its exploration spanning several decades and missions. Early reconnaissance began with Pioneer 10 (1973) and Pioneer 11 (1974), followed by the groundbreaking flybys of Voyager 1 and Voyager 2 in 1979, which unveiled Io's intense volcanism and dynamically reshaped surface. The Galileo orbiter (1995–2003) then provided comprehensive studies of Io's geology, tidal heating processes, and interactions with Jupiter's formidable magnetosphere. Additional insights were gathered during Cassini's Jupiter-flyby in 2000 and New Horizons' brief encounter in 2007. Currently, the Juno spacecraft, orbiting Jupiter since 2016, continues to refine our understanding of the Jovian system. Looking forward, upcoming missions such as NASA's Europa Clipper—launched in October 2024—and ESA's Juice (JUpiter ICy moons Explorer), launched in April 2023 with an expected arrival at Jupiter in 2031, promise to further illuminate the complex dynamics of Jupiter's environment, thereby providing additional, albeit indirect, insights into Io's evolution within this intricate system.

### III. Electrodynamic Tether Spacecraft Propulsion

Electrodynamic tether (EDT) propulsion is a propellantless maneuvering technique that leverages the interaction between a long, electrically conductive cable and a planetary magnetic field[7]. In the context of the UNAGI mission, this method is employed to achieve fine trajectory corrections and orbital synchronization with Io. As the EDT moves through a planetary magnetic field—Jupiter's, in the case of UNAGI—a change in magnetic flux is experienced, inducing an electromotive force (EMF) along the tether according to Faraday's law. The induced voltage is given by:

$$\varepsilon = vBL\sin\theta \tag{1}$$

For the induced EMF to drive a current, a closed electrical circuit must be established. This is typically achieved by integrating electron collection devices (e.g., plasma contactors) and electron emission mechanisms (such as cathodes). Once current flows through the tether, the interaction with the magnetic field produces a Lorentz force:

$$F = \frac{v\pi r^2 B^2 L \sin^2\theta}{\eta} \tag{2}$$

This force is oriented perpendicular to both the current and the magnetic field. By controlling the current's magnitude and direction, the tether system can generate thrust or drag, allowing the spacecraft to adjust its velocity without traditional propellant. By actively managing the electron collection and emission processes, the system controls the current such that the resultant Lorentz force provides the necessary thrust. For the UNAGI mission, this controlled thrust enables the spacecraft to "catch up" to Io by matching its orbital velocity, analogous to a surfer riding an incoming wave.

The UNAGI mission utilizes EDT propulsion to achieve its key objective—a controlled landing on Io. By leveraging the naturally occurring Jovian magnetic field[8], the system allows for dynamic, propellantless trajectory corrections. This capability is critical in a high-radiation, high-gravity environment where traditional chemical propulsion poses significant challenges. In UNAGI's design, precise modulation of tether current and orientation ensures that the spacecraft synchronizes its arrival with Io, enabling a safe and controlled descent while optimizing fuel efficiency and overall mission cost-effectiveness.

### IV. Mission Architecture

This mission design has multiple configurations that can accommodate different launch providers. If this mission is awarded a scope similar to that of a current New Frontiers budgetary class, the spacecraft is launched on an interplanetary trajectory toward Jupiter on a New Glenn launch vehicle[9]. If deemed competitive, this mission could be scoped to fit within the parameters of a secondary payload for a larger (New Frontiers, Flagship) mission that is bound for Jupiter. If neither of the preceding options are available, this mission could be launched to Low Earth Orbit (LEO) and can be pushed to an interplanetary trajectory utilizing a commercial solar electric propulsion solution. During the cruise phase, only minimal attitude corrections and periodic communication checks are required. At this stage, the deployment of solar arrays—totaling 150 m² of photovoltaic surface yielding 1070 watts[10] of electric power which is sufficient to support onboard electronics, communications, and later tether operations.

Upon arrival at Jupiter, a propulsion maneuver inserts the spacecraft into a carefully chosen elliptical orbit:

| Perijove (Jupiter Radius') | Apojove (Jupiter Radius') | Eccentricity | Semi-Major Axis | Inclination (degrees) |
|---|---|---|---|---|
| $8R_j$ | $20R_j$ | 0.43 | $14R_j$ | 60° |

The orbit is designed to intersect Io's orbital plane (with a mean orbital radius of ~421,700 km), thereby providing multiple opportunities for tether-based maneuvers. Once a stable Jupiter orbit is achieved, the spacecraft shall deploy a 50km electrodynamic tether from opposing faces of its dodecahedron-style bus.

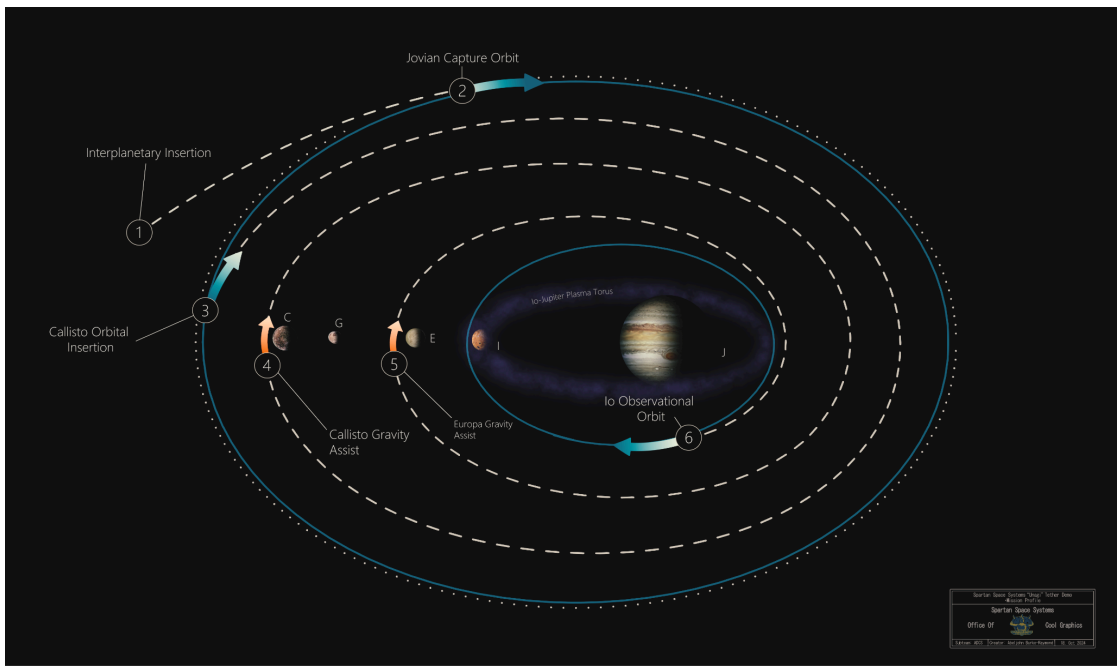

**Figure 1. Orbit Insertion Maneuver.**

The tether is fabricated from a high–strength aluminum alloy and engineered to achieve a mass per unit length of about 8.5[11]g/m For a 50km tether, this amounts to roughly 425kg—a fraction of the overall spacecraft mass. Deployment is achieved gradually using controlled reels mounted to isolate dynamic loads from the main bus, ensuring proper alignment with Jupiter's magnetic field, and minimizing permutations to the instrument and communications systems that require precision.

Using continuous, low–thrust maneuvers, the tether system gradually modifies the spacecraft's orbital parameters (semi-major axis, eccentricity, and phase) according to Gauss's planetary equations. This controlled modulation allows the spacecraft's orbit to "catch up" with Io. Solar array power is not only used to sustain the primary spacecraft systems but also to control the tether's current and associated charge–exchange systems that generate thrust. Over successive orbits, the spacecraft's trajectory is refined until it becomes nearly co-orbital with Io, with the relative velocity reduced to near zero at a precisely calculated encounter point.

High–precision navigation—integrating star trackers, inertial measurement units, and optical sensors—ensures accurate alignment of the tether thrust vector. In the final approach, the tether's control system fine-tunes the trajectory so that a relative-descent begins with minimal kinetic energy relative to Io's surface.

The spacecraft will simultaneously work to match Io's orbital velocity around Jupiter and its rotational velocity about its own frame of reference. The spacecraft's trajectory can be modeled as such:

$$u_1 \bullet = \frac{u_2^2}{r} - \frac{\mu}{r^2} + \frac{F_{L,r}}{m} \qquad (3)$$

$$u_2 \bullet = \frac{2u_1 u_2}{r} + \frac{F_{L,\theta}}{m} \qquad (4)$$

The spacecraft will pursue Io until it enters the moons' Sphere of Influence (SOI). From there the spacecraft will begin to feel a force of gravity from Io being exerted on it. The tether and gravitational pull from Jupiter and its moons' will work to reduce the impact of this force until other landing strategies are employed. Landing Architectures being evaluated right now include airbags deployed from an onboard solid catalyst that will sublimate upon activation, a springboard style dampener, and intentional crumpling zones of the spacecraft bus.

Once on Io, the spacecraft's internal instrument suite—housed within the radiation–shielded dodecahedron bus—becomes operational. A dedicated, deployable aperture allows optical and other remote–sensing instruments to monitor Io's surface and volcanic activity. Given Io's extreme environment, the bus features robust thermal control and radiation shielding, while the solar arrays (if exposed) continue to provide power during surface operations. Communications with Earth are maintained via a high–gain antenna, ensuring that critical science data are relayed throughout the operational phase.

## V.   Science

The UNAGI mission's science payload is designed to address high-priority objectives related to Io's geophysical, geochemical, and magnetospheric properties. These objectives include:

1) Investigating Io's internal structure and probing for a subsurface magma body.
2) Mapping surface modifications resulting from volcanic and hydrothermal processes.
3) Characterizing active melt generation, outgassing, and plume activity.
4) Assessing tectonic deformation and global-scale asymmetries.
5) Understanding interactions between Io's volcanic emissions and Jupiter's magnetosphere.
6) Comparing Io's processes with those observed on Europa, Enceladus, and other icy satellites.

Each science goal is linked with its measurement objectives, instrument requirements, and expected data products, ensuring that all observations directly contribute to answering key questions outlined in NASA's Decadal Surveys[12].

**A. Long-Focal-Length Cameras**
1) Configuration: Fixed, high-resolution cameras—similar in design to the Europa Imaging System—are mounted directly on the spacecraft bus.
2) Capabilities: The suite includes both wide-angle and narrow-angle configurations to achieve resolutions as fine as ~100 m per pixel during distant flybys, improving up to 100x in close approach.

**B. Camera Positioning**
1) Attitude Control: Advanced Attitude, Determination, and Control Systems (ADCS) analysis ensures optimal orientation during flybys, avoiding occlusion from the bus or other structures.

**C. Robotic Arm with Panoramic Camera**
1) Design**:** A ~1.8 m robotic arm—comparable in scale to that used on the InSight mission—is integrated to support close-up surface imaging. It can also provide spacecraft health and welfare imagery at key junctures.
2) Field-of-View: It carries a 360° panoramic camera, featuring a controlled 45° instantaneous field-of-view steered to cover the full sphere.

3) Stowage: The arm is designed to be non-protrusive during cruise and flybys, stowing securely against the bus when not in use.

**D. Magnetometer**
1) Deployment: Mounted on an extended boom, the magnetometer measures Io's magnetic environment to infer tidal and volcanic influences.

**E. Plasma Sensor (PIMS/CFIDS)**
1) Function: This instrument characterizes the local plasma environment, assessing Io's contribution to Jupiter's magnetospheric dynamics by measuring plasma temperatures and energetic electron populations.

**F. Seismometer**
1) Integration: A seismometer, similar in design to the InSight instrument, is included either on the main bus structure. It provides critical data on Io's internal structure and the depth and composition of any magma bodies. This instrument will be mounted opposite of the robotic arm relative to the spacecraft bus.

**G. Infrared (IR) Spectrometer and Gas Chromatograph**
1) IR Spectrometer: Maps thermal anomalies and analyzes the composition of volcanic plumes and surface materials.
2) Gas Chromatograph: Measures outgassing and volatile compositions to elucidate Io's atmospheric dynamics and the chemical nature of its volcanic outputs.

**H. Flyby and Landing Operations**
1) Imaging and Magnetic Measurements: The majority of high-resolution imaging and magnetic field data will be gathered during multiple flybys, enabling stereoscopic imaging and detailed geological mapping.
2) Lander Module Alternative: In the event a full spacecraft landing on Io is not feasible, an alternative configuration includes detaching a small lander module. This module would carry a subset of instruments (e.g., a high-resolution camera and seismometer) to perform localized surface investigations, thereby maximizing overall science return.

The integration of these instruments with the spacecraft's advanced GNC and EDT propulsion systems ensures that every measurement is coordinated with the dynamic mission profile, from cruise through landing, providing a comprehensive understanding of Io's complex and extreme environment.

## VI. Spacecraft Configuration

The central spacecraft bus is designed as a regular dodecahedron—a polyhedron with 12 pentagonal faces. This configuration offers several advantages. Twelve faces allow for flexible placement of instruments, antennas, and other subsystems. The nearly isotropic mass distribution simplifies attitude control and helps mitigate off-nominal torques arising from tether forces or solar radiation pressure. With overall dimensions of approximately 2–3 m per side (totaling about 10–15 m³), the interior is optimized to house science instruments, avionics, power management, and thermal control components.

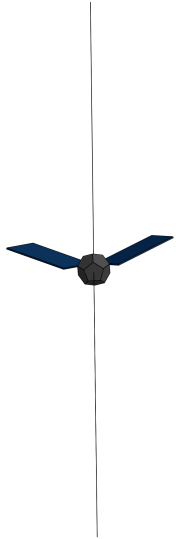

**Figure 2. UNAGI Model.**

  The spacecraft's payload bay is housed inside the bus in a radiation vault style drawing heritage to recent flight missions to Jupiter. It will leverage the pentagon shaped faces of our bus to create a hinged door similar to that of Hubble Space Telescope[13] or its KH-11 bus design so that our payload is protected from the harshest of the Jovian radiation environment at key times and as well as landing.

  This approach facilitates passive thermal control while protecting critical components from intense radiation near Jupiter and Io. Additionally, two opposing faces are dedicated to tether attachment; the tether is mounted on extendable booms or articulated arms to isolate dynamic loads, thus minimizing oscillations or bending stresses during deployment and operation. The electrodynamic tether is a high-leverage, lightweight element that plays a central role in the mission's propulsion strategy. The spacecraft employs two tethers each at 25 km on opposing ends of the spacecraft bus, affixed with a spool system. This is sufficient to generate the necessary Lorentz force in Jupiter's magnetic field. Constructed from a 6061 aluminum alloy designed for radiation resistance and micrometeoroid protection. Targeting a mass per unit length 8.5g/m results in a total tether mass of roughly 425kg. The tether is coiled on a motorized reel system and deployed gradually following stable Jupiter orbit insertion. Deployment control relies on onboard sensors and guidance, navigation, and control (GNC) systems to maintain optimal alignment with the local magnetic field. Integrated electronics regulate the tether current, adjusting both thrust magnitude and direction, with the GNC algorithms enabling gradual orbital phase adjustments with Io.

  Solar panels are strategically arranged on the exterior of the bus, between the tether mounting points, as large deployable wings. These arrays can articulate to face the Sun optimally, ensuring continuous exposure despite attitude changes. At Jupiter's distance, the reduced solar flux is approximately 50–55 W/m². Assuming a ~20% efficiency, about 10–11 W/m² is available. For a mission power requirement of roughly 1KW (covering avionics, tether control, communications, etc.), the total solar array area is approximately 150 m$^2$. The arrays use lightweight, radiation-hardened cells with robust articulation mechanisms to support continuous power generation and tether operations.

  The spacecraft is equipped with robust, radiation-hardened processors and solid-state data storage to securely handle the science data from Io's surface operations. A suite of star trackers, inertial measurement units, and optical navigation sensors provides high-precision attitude and position data. Reaction wheels and thrusters are employed for fine orientation adjustments, ensuring that the designated landing face or viewing aperture is accurately aligned. Communication is facilitated through a deployable high-gain antenna—either integrated into one of the dodecahedron faces or as an extendable boom—that supports high-bandwidth data transmission during both cruise and surface operations. These ensure continuous contact during critical phases, such as tether deployment and the Io landing sequence. The bus construction employs 6061 aluminum alloy with strategic reinforcements at key locations (e.g., tether mounting points, the viewing aperture, and edges susceptible to impact loads). The designated landing face is reinforced and may include deployable shock absorbers to reduce impact forces during touchdown. Additionally, multi-layer insulation, passive heat pipes, and radiators are integrated to manage and dissipate heat generated by onboard electronics and tether control systems. The tether's design, featuring a multi-strand configuration and carefully selected materials, ensures resilience against micrometeoroid impacts and long-term degradation in the harsh Jovian environment.

Resilient components will be especially critical for the telecommunications aspect of the mission as well. The high radio emissions of the Jovian system will require precise modeling and estimation of power transmitted. At most, the spacecraft will be equipped with a 3m wide high gain antenna with a low-gain backup antenna. The geometry and landing trajectory of the spacecraft also require unique geometry and placement of the antenna. Mitigation of landing damage and impact are key. Additionally, the ground system will have to be provided through the NASA Deep Space Network (DSN). This network consists of three, 34m diameter, 20kW antenna arrays. These arrays are easily powerful enough to reach the Jovian system, however, in the event of emergencies there are two 80kW, high power arrays. In terms of data storage and bitrate, bitrates between 100-1000 kbps are the current estimate, taking into consideration that our likey modulation scheme is Quadrature Phase-Shift Keying (QPSK). Despite the heavy noise, QPSK has much higher data rates and can be viable with proper error correction code.

## VII. Discussion

Within the context of deep space and even near earth missions, propellantless propulsion is necessary and important for several reasons. Using an EDT maximizes fuel economy by leveraging the environment our spacecraft operates in. This can free up mass allocations to boost payload or other subsystems. An EDT system can also reap benefits through longevity. If satellites or probes never need to refuel they can theoretically stay in orbit for much longer, potentially indefinitely. All of these useful attributes ultimately contribute to the most useful thing about the EDT system: cost. Due to an EDT having no propellant, and no chemical propulsion system, the complexity of manufacturing decreases and thus costs also decrease. If the EDT is able to remain in orbit for longer, less satellites and probes need to be replaced which also reduces costs. It's extremely likely that if an EDT system is perfected, it will significantly decrease the cost to manufacture and launch satellites not only for LEO but also for deep space use.

Earth, Jupiter, Saturn, Uranus, and Neptune all feature palpable magnetospheres. If this technology is leveraged correctly, it can lead to a boon in planetary exploration. Leveraging this technology in the vicinity of Earth can lead to commercial and earth science economic benefits that affect such spacecraft systems as deorbit technology, stationkeeping, and attitude adjustments. For the rest of the solar system, the gas and ice giants environments host plasma that can be surfed on to reach far flung moons and orbital longevities far exceeding conventional chemical or even noble gas propelled missions.

## VIII. Conclusion

This team seeks to develop our concept further using detailed simulation tools such as Systems Toolkit and detailed spatial models of Jupiter's magnetic field. Such a model is maintained by NASA's Jet Propulsion Laboratory as well as its existing Solar System Ephemeris. These datasets will aid us in honing in our trajectory that accounts for charge distributions, small bodies, and precise dynamics. The team seeks to employ finite element simulation tools that will be able to specify design parameters for structures like the tether, the spacecraft bus, mounting devices, and mechanisms of the space vehicle.

This team also is working with its institution to begin a campaign of lab testing to characterize tether behavior in a near-space environment. This will include mechanics of materials testing of different candidate tether materials, vacuum chamber testing of current induction in varying magnetic fields, and possibly high altitude balloon based flight tests in the future.

## Acknowledgements

This team acknowledges the expertise and steadfast wisdom and service of Geologist Suzy Brookshire in contributing to the pool of knowledge this mission has in the realm of planetary geology. Her professionalism and capability cannot be overstated and she tremendously aided this team in its pursuit of delivering a capable spacecraft proposal.